# Uranium partitioning between liquid iron and silicate melt at high pressures: implications for uranium solubility in planetary cores

Xuezhao Bao[1*], Richard A. Secco[1], Joel E. Gagnon[2], Brian J. Fryer[2]

[1]Department of Earth Sciences, University of Western Ontario, London, Ontario, Canada N6A 5B7

[2]Department of Earth Sciences, University of Windsor, Windsor, Ontario, Canada N9B 3P4

**Abstract**

We have investigated the partitioning of U between silicate melt and Fe liquid at pressures of 3.0 to 14.5 GPa and temperatures of 1660 to 2500 $^o$C. The solubility of U in liquid Fe is in the range of 0.6 to 800 ppm and increases with temperature (T) and pressure (P). When P $\geq$ 7 GPa and T > $T_{melt}$ of the silicate phase (olivine), the U concentration in Fe is 3 to 5 times greater than for run products where T < $T_{melt}$ of the silicate phase. Correspondingly, partitioning coefficient $D_U$ values can reach 0.031 at 8.5 GPa (using BN sample container) and 0.036 at 14.5 GPa (using graphite sample container). This implies that if a terrestrial-type planetary core with a pure Fe composition formed from a deep magma ocean (T > $T_{melt}$ of the silicate phase), then > 2.4 ppb U could have entered the core. Alternatively, if a core with same composition formed by percolation (T < $T_{melt}$ of the silicate phase), then based on the experimental results indicating $D_U$ increases with increasing P, 1 to 4 ppb U may have entered the core. The concentration of Si in liquid Fe also increases with pressure. The concentration of Ca is < 0.44 wt % for most samples and no relation with U concentration is found, which indicates that U may alloy with Fe directly. If Si concentration in the Fe phase can be used as an indicator of oxygen fugacity, then the increase in Si and U with pressure suggests a pressure dependent decrease in oxygen fugacity. This supports U (and possibly also Si) inclusion in the Earth's core at the time of core formation. The implications for radioactive heating in the planetary cores are briefly discussed.

---

[*] Corresponding author. 1-519-661-3157; Fax: 1-519-661-3198.
E-mail addresses: xuezhaobao@hotmail.com (Xuezhao Bao), secco@uwo.ca (Richard A. Secco)

## 1. INTRODUCTION

Uranium, thorium and potassium are the main heat-producing elements in the Earth. Among these, U and Th are the most important because their heat contribution is estimated to be approximately 85% of the total radioactive heat produced (Verhoogen, 1980). All three elements are lithophile at low P and T (Righter, 2003), and are thought to reside mainly in the continental crust. However, recent core energy models have proposed the possibility of these radioactive elements being an important energy source in the core (Labrosse et al., 2001; Buffett 2002; 2003; Nimmo et al., 2004).

The experimental results of solubility of K in liquid Fe at high pressure (Chabot and Drake, 1999; Gessmann and Wood, 2002; Murthy et al., 2003) suggest that $^{40}$K is a possible radioactive source in planetary cores. However, McDonough (2003) argued that the depletion pattern in the bulk Earth compared to meteorites for the moderately volatile lithophile elements (including K) does not favor the sequestration of K in the core. It is noteworthy that all of the experimental work thus far that observed significant K solubility in the Fe phase was achieved in high S starting materials (Chabot and Drake, 1999; Gessmann and Wood, 2002; Murthy et al., 2003). When S is less than 2.5wt% in the metallic phase, Chabot and Drake (1999) did not detect K in the metallic phase. If Si is the main light element in the core, and S is less than 1.7% (Dreibus and Palme, 1996) or 2.3% (Kilburn and Wood, 1997), then on the basis of the above mentioned experimental solubility results, only a minor amount of K should be expected in the core. Hence, Chabot and Drake (1999) argued for less than 1 ppm K in the core, and which would provide power 2 to 3 orders of magnitude lower than the estimates of the power necessary to drive the Earth's dynamo. Gessmann and Wood (2002) also concluded that

[40]K can only offer at most 20% of the total heat production of the core.

There are no data on the partitioning of U and Th between silicate and Fe phases under the conditions relevant to core formation. Murrell and Burnett (1986) performed experiments on the solubility of U and Th in FeS at low pressure (≤1.5 GPa). Their work showed that U may combine with CaS in the FeS phase. Later, McDonough (2003) argued that if U entered the Earth's core with Ca, then the current Ca/Ti and Ca/Al ratios of the mantle would be difficult to explain. The question of U and Th solubility in liquid Fe and their potential for supplying heat to the core is, therefore, still open and is pursued in this study.

## 2. EXPERIMENTAL AND ANALYTICAL DETAILS

### 2.1 The starting material

The starting material in each experiment was produced by mixing and grinding Fe (Aldrich Chem. Co., 99.99% pure), natural peridotite (containing 60-80 wt% Mg-olivine, from Bay of Islands ophiolite suite, Newfoundland, Canada) and uraninite (containing approximately 80 wt% $UO_2$, Goldfields, Saskatchewan, Canada). The mixture was composed of 40 wt% metallic Fe, approximately 54 wt % silicate and oxides and approximately 6 wt% $UO_2$. A sufficient amount of $UO_2$ was added to the sample in order to create conditions whereby the concentration of U in the Fe phase might exceed the detection limit of standard analytical instruments.

### 2. 2 High pressure and temperature experiments

The high-pressure experiments were carried out in a Walker-type module high-pressure device (Walker et al., 1990). The pressure calibration of the press was described in previous work (Secco et al., 2001). The pressure cell was a cast, pre-gasketed, MgO octahedron with edge length of 18, 16 and 14 mm for tungsten carbide cubes with truncations of 8, 6, or 4 mm, respectively. Approximately 40 to 50 mg of sample powder was loaded into a boron nitride (BN) or graphite capsule, which occupied the center of the pressure cell. The capsule was surrounded by a cylindrical Nb or graphite furnace, which was thermally insulated from the MgO octahedron by a zirconia sleeve. An MgO plug and sleeve were used to separate the sample capsule and Fe conduction rings at the ends. The temperature was measured by a W-3%Re/W-25%Re thermocouple situated directly above the capsule and pressure correction for the emf was not applied. All assembly components were fired at 200 $^{o}$C for one hour before loading. For the lowest pressure 3GPa runs, pressure was slowly increased to the target pressure at a pressurization rate of approximately 2GPa/hr. For all other runs, pressure was increased rapidly to 2.5GPa followed by a pressurization rate of approximately 2GPa/hr to the target pressure. After being compressed to the desired pressure, the temperature was then raised at 100 $^{o}$C/min to the desired run temperature and then held there for 3 to 73 min (Table 1). Temperature gradients were measured in prior experiments, and values of 100$^{o}$C/mm were obtained at the maximum temperature in this study.

After completion of the experiment, the entire sample assembly was mounted in epoxy, sectioned through the center of the sample and then polished for Laser Ablation Inductively Coupled Plasma Mass Spectrometry (LA-ICP-MS), Electron Microprobe (EMP), and Secondary Ion Mass Spectrometry (SIMS) analyses.

2.3 Electron Microprobe

The major elements and light elements (C, N and O) in the silicate, metal and uraninite phases were analyzed by electron microprobes. For major element analyses, electron microprobes at the University of Western Ontario (Western), University of Manitoba (Manitoba), University of Toronto (Toronto), and Canada Centre for Mineral and Energy Technology (CANMET) at Ottawa (Ottawa) were used. The electron microprobe at Toronto was mainly used to analyze U in metal phases and the microprobe at CANMET was used for comparison since both laboratories have U standard materials. Similarly, uraninites were analyzed at Manitoba. For the light element analyses, the electron microprobe from McGill University was used.

Most analyses were conducted using a Jeol 8600 EMP at Western. An electron beam with a voltage of 15 kV and a current of 15 nA was used. Raw data were reduced using the ZAF correction built into Tracor Northern automation system. The beam size is 1 µm in this study. The depth of the electron beam penetration is 1 to 1.5 µm. Several samples were reanalyzed using other EMPs mentioned above at similar conditions. Usually for major elements, the accuracy of the microprobe is approximately ± 1.5 %.

The detection limits under the conditions used in this study were: Mg ~300 ppm; Si ~210 ppm; Ca 260 ppm; C, N and O: > 700 ppm.

2.4 Laser Ablation Inductively Coupled Plasma Mass Spectrometry

Analysis of U and other trace elements were conducted at the LA–ICP–MS facility in the Great Lakes Institute for Environmental Research at the University of Windsor using procedures and conditions described in Gagnon et al. (2003). With an optical system in a microscope, a laser beam is focused on the surface of a sample. The laser beam size can be adjusted from 10 to 40 μm in diameter according to exposure area. Two glasses from the National Institute of Standards and Technology (NIST), NIST 612 and NIST 610 were used as the external standard. Usually, the standards were run at both the beginning and end per phase every three experiments. Fe (for the Fe phase) and Mg (for the silicate phase) from the electron microprobe analysis were used as the internal standards. Every phase in each experiment was measured two to ten times depending on its exposure area. The measured data were reduced using the LAMTRACE program, written by Dr. Simon Jackson of MacQuarrie University, Australia. The detection limits are: U 0.01 ppm, Th 0.04 ppm; Mg 4.76 ppm and Ca 102 ppm.

2.5 Secondary Ion Mass Spectrometry

In this study, a SIMS (at Western) was also used for the U, Th, Ca, Ti, Si and Mg analysis in both metal and silicate phases, as a complementary analytical tool to LA-ICP-MS, because it can provide a profile of elemental concentration as a function of depth from the surface to deep interior. Although a standard material for quantitative analysis was not found, SIMS can qualitatively indicate if U has entered metal phase.

3. RESULTS AND DISCUSSION

3.1 P, T range of the experiments

The experimental conditions are given in Table 1. In all runs, the experimental temperatures were more than 200 $^\circ$C higher than the melting temperature of Fe, but most of them were lower than the melting temperature of Mg olivine (the main mineral (60-80%) in the silicate part of the starting materials) as shown in Fig.1. Fig.2 is a backscattered electron image of the recovered product of run 182, which shows Fe separated from the silicate phase. This further indicates that the experimental temperature was high enough to melt the Fe phase at the experimental pressures. Large crystals (>200 μm) of silicate phase (the grain size in the starting material was < 50 μm) indicates that the temperatures reached the re-crystallization temperature of the silicate phase.

**Table 1 Experimental conditions**

| Run No. | T ($^o$C) | P (GPa) | Time (min) | Sample Container |
|---------|-----------|---------|------------|------------------|
| 115 | 2000-2030 | 3 | 5 | Graphite |
| 200 | 1822-2050 | 3 | 3 | BN |
| 204 | 1731-2030 | 3.5 | 9 | BN |
| 205 | 1764-2021 | 5 | 11 | BN |
| 182 | 1886-2050 | 5 | 9 | BN |
| 183 | 1890-2050 | 6 | 4 | BN |
| 157 | 1790-2143 | 7 | 73 | Graphite |
| 201 | 1755-2076 | 7 | 17 | BN |
| 150 | 1888-2024 | 7 | 56 | Graphite |
| 190 | 1780-2212 | 7 | 7 | BN |
| 202 | 1859-2050 | 7.5 | 8 | BN |
| 212 | 1728-1996 | 8 | 42 | BN |
| 199 | 1760-2249 | 8 | 8 | BN |
| 191 | 1882-2288 | 8.5 | 3 | BN |
| 153 | 1665 | 8.5 | 1 | Graphite |
| 154 | 1887-2018 | 8.8 | 17 | Graphite |
| 214 | 1732-2077 | 9 | 12 | BN |
| 149 | 1871-2200 | 9.4 | 12 | Graphite |
| 253 | >2200 | 12 | 18 | BN |
| 167 | 2350-2485 | 14.5 | 60 | Graphite |

3.2 Si in Fe phase

The major components and possible light elements in the metal phases are listed in Table 2.  Si is one of these components. According to Hume-Rothery (1966), Si can form a solid solution with Fe. Since no metallic Si was added to the starting material, the

Si found in the recovered Fe phase most likely originated from the following reaction (Siebert et al., 2004):

$$SiO_2 \text{ (silicate)} + 2Fe \text{ (metal)} \leftrightarrow 2FeO \text{ (silicate or metal)} + Si \text{ (metal)} \quad (1)$$

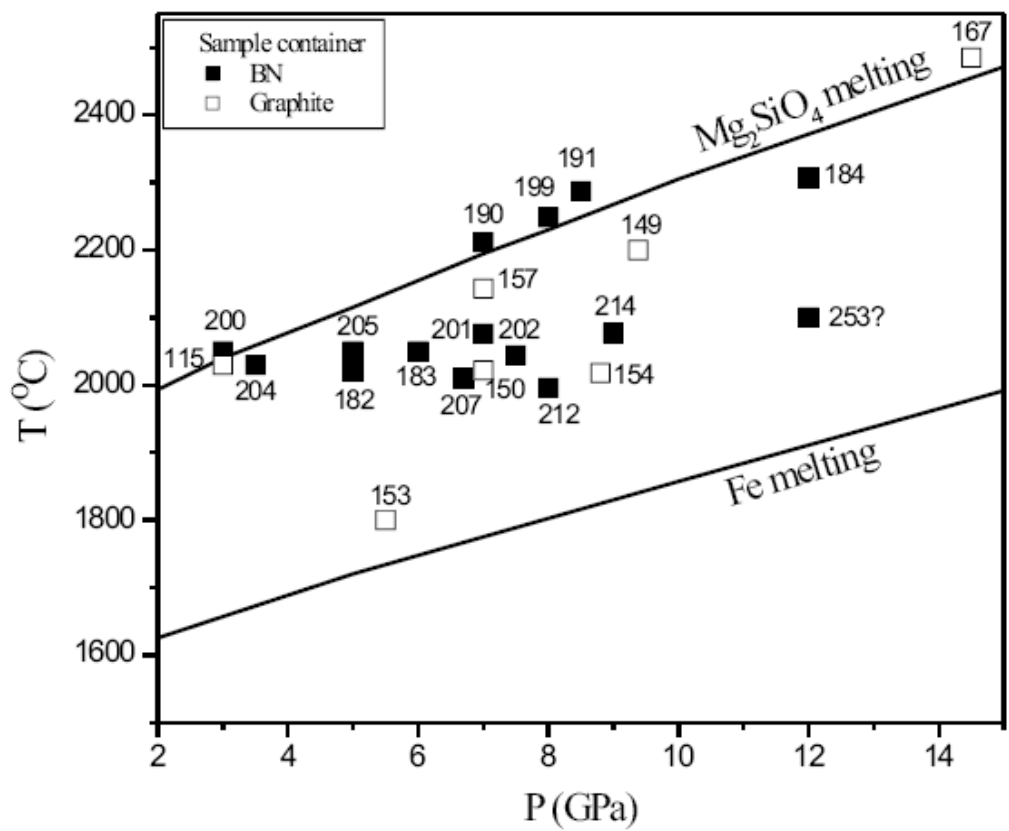

**Fig.1 The P and T conditions of the experiments. The temperature values are the highest experimental T in each run as measured by thermocouples (except run 199, which was determined from the correlation between furnace power and T). The T of run 253 is uncertain since a correlation between thermocouple power and T has not been completely established in the 4mm pressure cell (This value is estimated from its power value).**
**The reference olivine melting line is based on the data of Davis and England (1964), and Ohtani and Kumazawa (1981); and Fe melting line was based on the data of Saxena and Dubrovinsky, 2000.**

The data from EMP analysis revealed that there is 0.1 wt% to 4.0 wt% (log values: −1.62 to 0.58) Si in the Fe phases of the run products as shown in Fig.3. This is discussed in detail in the following sections.

### 3.2.1 Pressure influence

The sensitivity of LA-ICP-MS was adjusted to measure U with a concentration of 1 to 1,000 ppm in the Fe phase after several tests. Because the concentration range of Si (1,000 to 40,000 ppm) is much higher than that of U in the Fe phase, the EMP is more dependable for Si measurement. Hence, only EMP was used for Si analysis. Fig.3 clearly indicates that log wt% Si in the Fe phase from both graphite and BN containers increases

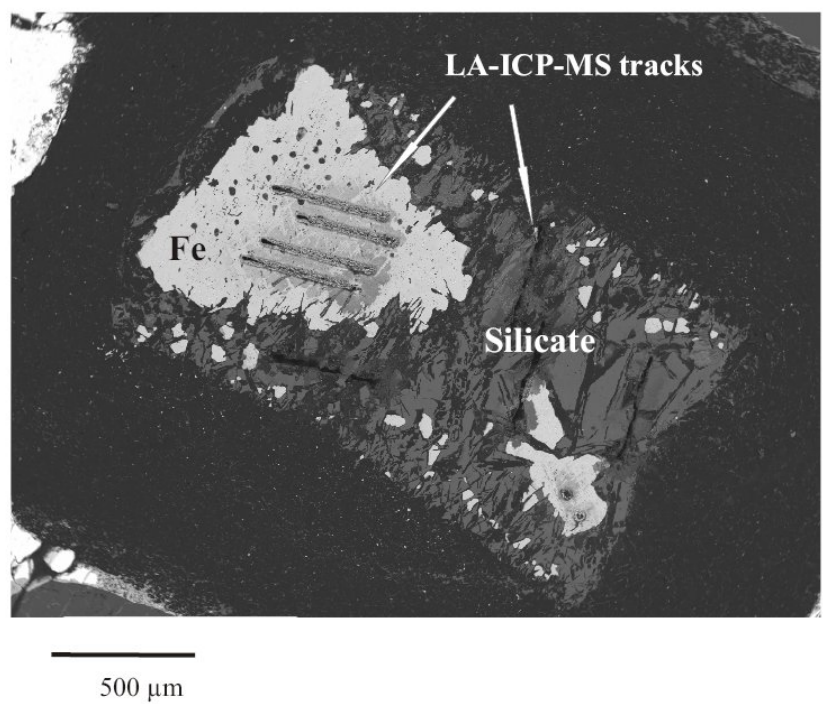

500 µm

**Fig.2 Recovered sample from run 182 showing the texture of the Fe (white areas) and silicate (grey areas) phases. The laser beam tracks of LA-ICP-MS sampling can be seen in both phases**.

with pressure. This is consistent with previous studies (Ito et al., 1995; Gessmann et al., 2001; Malavergne et al., 2004).

According to Kilburn and Wood (1997), the concentration of Si in the Fe phase can be used as an indicator of oxygen fugacity, $fo_2$. From Fig.3, it can be seen that from 3 GPa to 14.5GPa, Si in the Fe phase increases from $-1.62$ log unit to 0.58 log unit. When compared with the results of Kilburn and Wood (1997), this corresponds to a decrease in oxygen fugacity from $\log f_{O2} = -2$ to $\log f_{O2} = -6$ below the iron-wüstite (IW) buffer. This means oxygen fugacity decreases with increasing pressure, or oxygen concentration in the Fe phase decreases with increasing pressure. This is consistent with the conclusion of Rubie et al. (2004), who showed that increasing pressure decreases the solubility of oxygen in the Fe phase. Therefore, our experimental results indicate that pressure can create low oxygen fugacity or a reducing environment.

Another point worth noting is that although the Si content of the Fe phases in both BN and graphite containers increases with pressure, the band of data for BN container trends with a larger slope than the band of data for graphite container. Gessmann and

**Table 2 Average composition in the Fe phase analyzed by Electron Microprobe (wt%)**

| run # | Fe | S | Si | Mg | O | B | N | Total** |
|-------|------|------|------|------|------|------|------|---------|
| 115 | 96.38 | 0.05 | 0.01 | 0.00 | | | | 96.44 |
| 149 | 92.95 | 0.10 | 0.00 | 0.01 | | | | 93.06 |
| 150 | 85.97 | 0.04 | 0.18 | N/A | | | | 86.31 |
| 153 | 81.20 | 0.12 | 0.55 | N/A | | | | 82.00 |
| 154 | 89.12 | 0.08 | 0.19 | N/A | | | | 89.45 |
| 157 | 85.28 | 0.93 | 0.09 | 0.00 | | | | 86.30 |
| 167 | 84.74 | 0.08 | 3.11 | 0.15 | | | | 88.08 |
| 182 | 99.07 | N/A | 0.23 | 0.00 | | | | 99.98 |
| 183 | 96.86 | N/A | 0.06 | 0.00 | | | | 97.91 |
| 184 | 87.12 | N/A | 2.26 | 0.01 | | | | 90.84 |
| 190 | 94.59 | 0.01 | 0.31 | 0.00 | | | | 94.91 |
| 191* | 94.12 | 0.03 | 0.51 | 0.00 | 0.88 | 0.08 | 1.26 | 96.88 |
| 199 | 93.39 | 0.02 | 0.40 | 0.00 | | | | 93.82 |
| 200* | 89.08 | 0.02 | 0.00 | 0.00 | 0.65 | 0.14 | 2.23 | 92.12 |
| 201 | 96.73 | 0.00 | 2.22 | N/A | | | | 99.60 |
| 202* | 94.86 | 0.01 | 2.49 | 0.00 | 0.90 | 0.09 | 0 | 100.05 |
| 204 | 92.48 | 0.26 | 0.25 | 0.14 | | | | 93.67 |
| 205 | 84.84 | 0.02 | 0.71 | 0.09 | | | | 85.67 |
| 207 | 96.31 | 0.02 | 0.45 | N/A | | | | 97.17 |
| 212 | 88.02 | 0.73 | 2.68 | 0.01 | | | | 91.42 |
| 214 | 82.74 | 0.00 | 2.27 | 0.07 | | | | 85.08 |
| 226 | 90.66 | 0.38 | 0.01 | 0.00 | | | | 91.05 |
| 253 | 90.10 | 0.28 | 2.59 | 0.12 | | | | 93.08 |

**\*   Runs with light element analyses.**
**\*\* The metal phase may contain other elements, such as light elements B, N and O coming from container and starting materials as shown in run 191, 200 and 202.**

Wood (2002) reported that the K partitioning coefficient of Fe sulfide samples in graphite containers was lower than those in MgO and $Al_2O_3$ containers. They concluded that carbon (C) coming from the graphite containers was the cause, and suggested C can migrate from the container to the Fe phase during high P, T experiments, which lowers the solution of K in the Fe phase. Other data also show that up to 8% C can diffuse into the Fe phase and form an interstitial solid solution (Hume-Rothery 1966). Noting that C and Si come from the same group in the periodic table, it is not surprising that C can partially substitute for Si in the Fe phase, which lowers the Si uptake of Fe phases from samples run in graphite containers. Although B (< 0.4 wt %) and N also can enter the Fe phase (Hume-Rothery 1966), as shown in Table 3, their chemical (valence) and physical properties are unlike those of Si. Therefore, B and N appear to have small competing influences on Si for residence in the Fe phase.

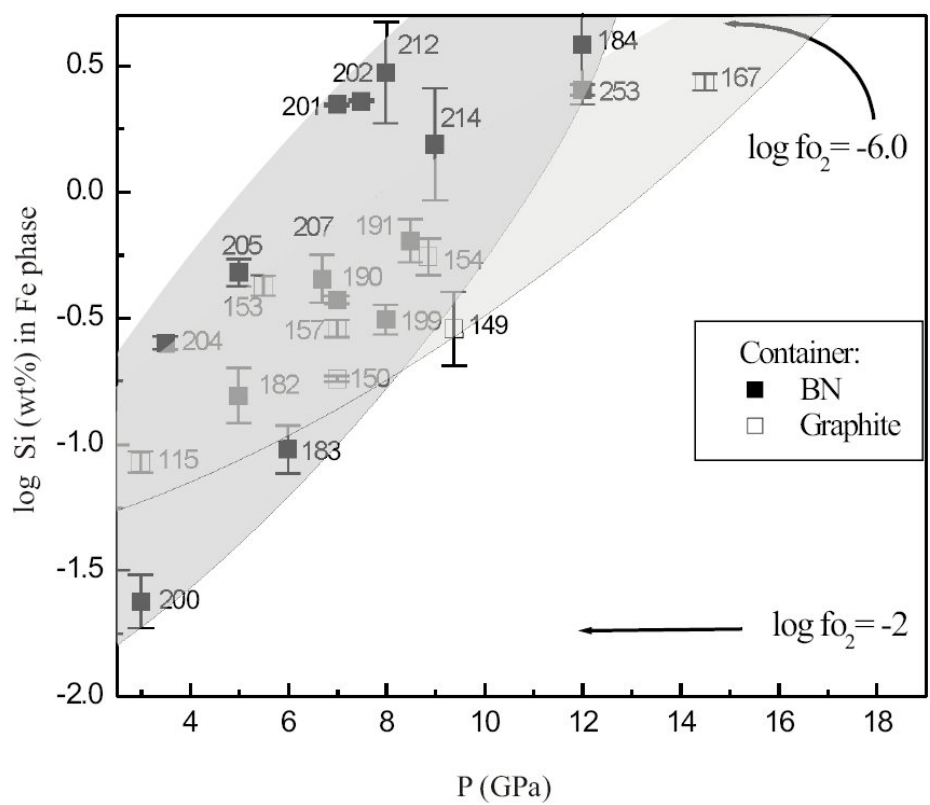

**Fig.3 Si in the Fe phase as a function of pressure. The shaded region with a lower trending slope is for runs with graphite containers, the other is for BN containers. Based on the method of Kilburn and Wood (1997), the oxygen fugacity range below iron-wustite (IW) buffer was calculated. Errors are 2σ.**

The effect of C vs. Si incorporation in Fe on the lattice cell volume may act as a control on U solubility in Fe. When C substitutes for Si, it will decrease the volume because the atomic radius ratio $r_C/r_{Fe}$=0.613, is much smaller than $r_{Si}/r_{Fe}$=0.94. This will not favor U substitution for Fe in the metal phase because the atomic radius of U ($r_U/r_{Fe}$=1.238) is larger than that of Fe. This may explain why the U concentration in products from graphite containers is generally lower than those from BN containers as shown in Fig.4.

3.2.2 Temperature influence

From Fig.3, it can be seen that the Si content of the Fe phase increases with pressure. But when pressure is fixed, increasing temperature leads to a decrease in the Si content of the Fe phase as shown in Fig.5. This behavior is consistent with experimental results from other studies (Geßmann and Rubie, 1998; Lin et al., 2002) as shown in Fig.6. A possible cause is that increasing temperature will lead to volume expansion of Fe lattice. Since the radius of Si is only 94% of that of Fe, an increase in the Fe-Fe distance may not favor the smaller Si to substitute for Fe. Therefore, temperature appears to have an opposite effect compared with pressure on Si content in the Fe phase.

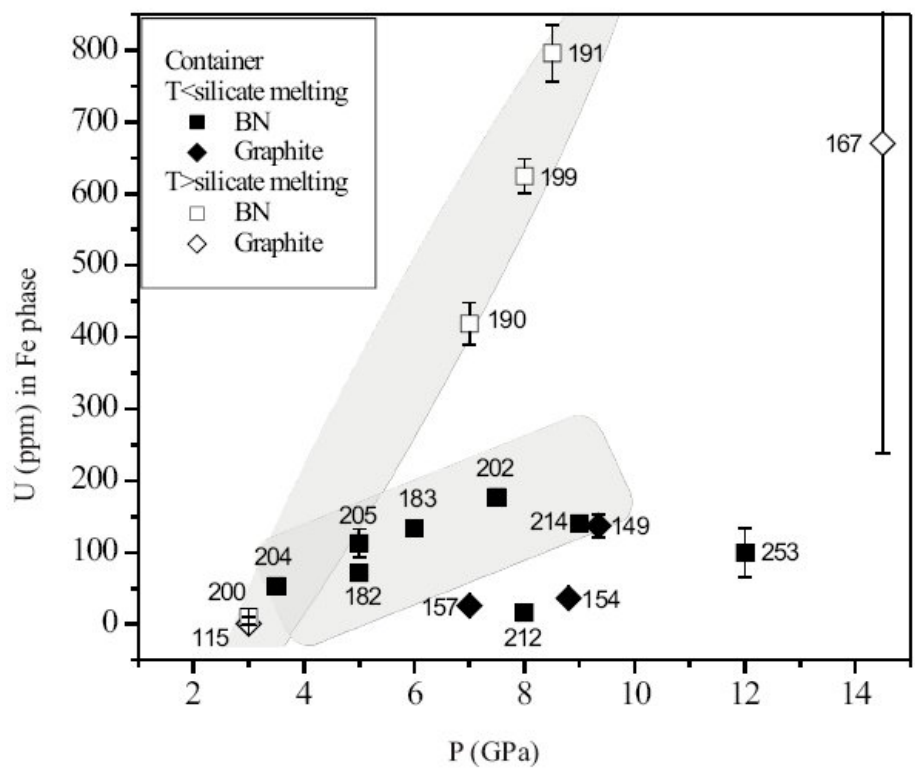

**Fig.4 U concentrations in the Fe phase as a function of pressure. According to the experimental T, they are divided into two main groups: T < $T_{melt}$ of silicate, the shaded region with a lower trending slope, and T > $T_{melt}$ of silicate, the shaded region with a larger trending slope. The Fe phases of some run products in Fig.1 are too small to satisfy the analysis requirement of LA-ICP-MS (>15 μm), therefore, no data came from these runs. Errors are 2σ.**

### 3.3 U in Fe and silicate phases

### 3.3.1 LA-ICP-MS and EMP analysis and results

Quantitative analysis was carried out by LA-ICP-MS and EMP. The detection limits for the LA-ICP-MS analytical method for the elements analyzed are: U 0.01 ppm; Th 0.04 ppm; Mg 4.76 ppm; Ca 102 ppm. Compared with the detection limit of U by EMP of 150-220 ppm, the data from LA-ICP-MS are much more quantitatively reliable. The U concentration of Fe phases of the run products from LA-ICP-MS is in the range of 1 to 1000 ppm, and is mostly distributed in the range 50–200 ppm. This is below the detection limit of EMP, therefore, the error in the EMP results is much larger than the error from the LA-ICP-MS analyses, which are presented in Table 3. The measured U values using EMP are in the range of 350-450ppm in the Fe phases of 4 run products (runs 182, 183, 184, 202) at different pressures. These are much higher than the corresponding U values from the same products from LA-ICP-MS as shown in Table 3. This may be explained by one of the following two reasons: a) The U concentration in the Fe crystal is too close to or below the detection limit (see Table 3), hence the EMP method is not sensitive enough to reflect the real U concentration in the Fe phases, b) there is overlap between U peak and Fe peak in the WDS spectrum of EMP. Therefore, the U and Ca data from LA-ICP- MS were used as the average concentration of samples.

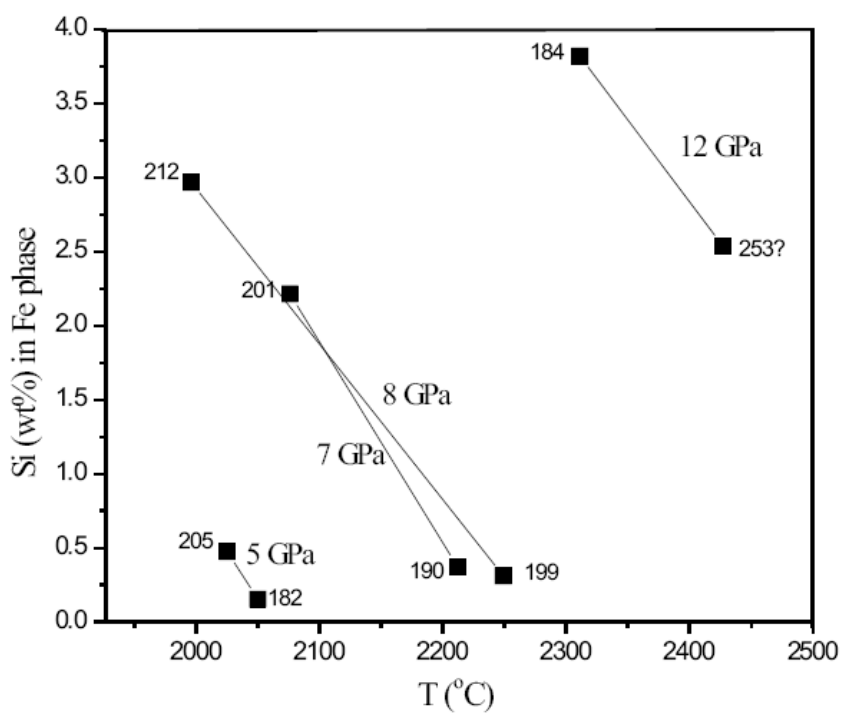

**Fig.5 Si content in the Fe phase with increasing temperature at fixed pressure. The sample containers in each run product are BN. The T in run 253 is uncertain as explained in Fig.1. Errors are 2σ.**

In the Fe phase, the sampling depth of LA-ICP-MS is in the range 25 to 30μm and the laser beam width is 10μm in most of the samples listed in Table 3. As shown in Fig.2,

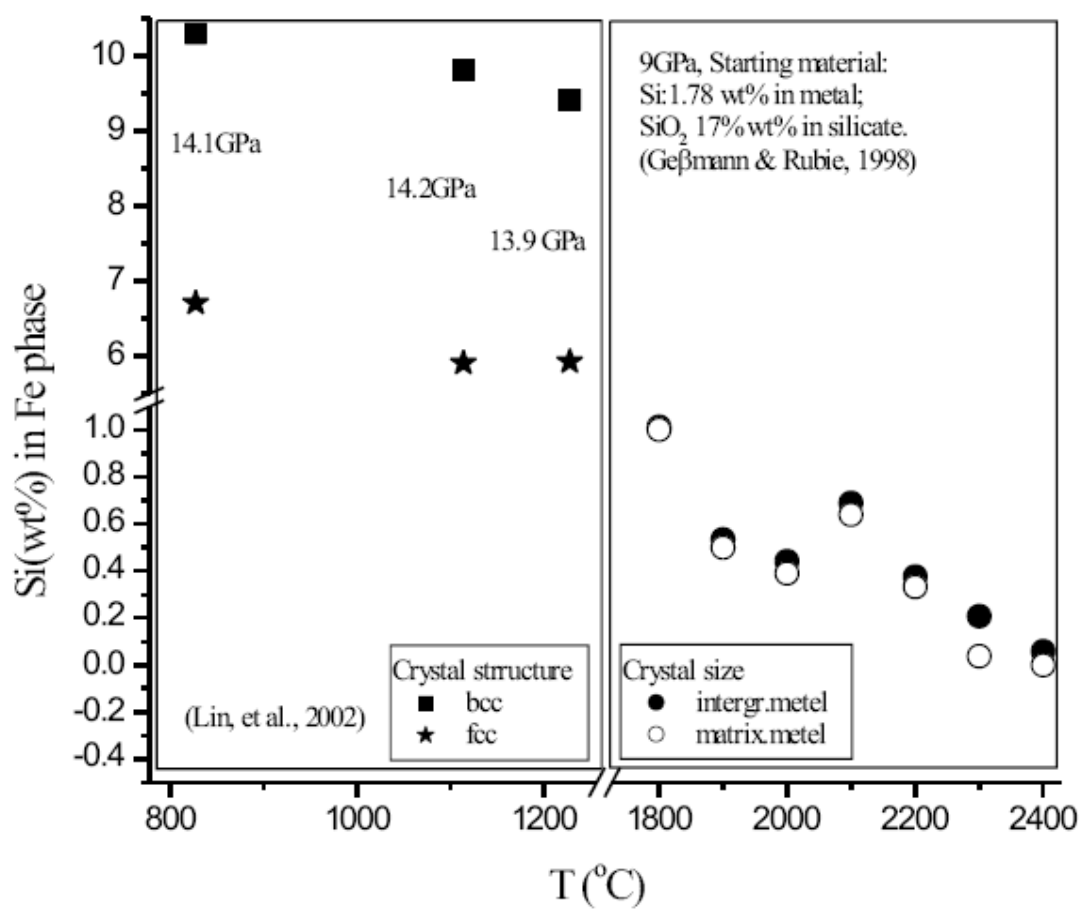

**Fig.6  Si content in Fe phase with increasing temperature at fixed pressure based on the data of Lin et al. (2002) and Geßmann and Rubie (1998).**

most of the laser sampling tracks is across large portions of the entire crystal. Hence, the sampling volume of LA-ICP-MS (hundreds of times larger than that of EMP) is large enough to represent the average composition of the sampled phase. But large sampling volume makes it possible to include small inclusions of the silicate phase in the analyses of the Fe phase, especially if the Fe phase is thin and underlain by silicate. In the silicate, Si and Mg are the major elements. As discussed above, Si is soluble in the Fe phase, but Mg has low solubility in Fe phases at high temperatures and pressures (Massalski, 1986; Ito et al., 1995). Therefore, Mg can be used as an indicator of whether or not the data include a contribution from the silicate phase. In EMP analysis, the electron beam is

**Table 3 Analysis results from LA-ICP-MS**

| Run No. | U in Fe phase (ppm) | U in silicate phase (ppm) | $D_U$ | Mg in Fe phase (wt%) | Ca in Fe phase (wt%) | Beam size (μm) | Container |
|---|---|---|---|---|---|---|---|
| 200 | 0.7±0.7 | 31277±199 | 0.00002±0.00002 | 0.000 | 0.039 | 10 | BN |
| 204 | 53± 3 | 33242±970 | 0.00160±0.00011 | 0.019 | 0.214 | 10 | BN |
| 205 | 112 ± 20 | 28735±498 | 0.00392±0.00071 | 0.059 | 0.447 | 10 | BN |
| 182 | 71 ± 0.1 | 28215±541 | 0.00252±0.00003 | 0.023 | 0.035 | 10 | BN |
| 183 | 133 ± 3 | 38012±338 | 0.00352±0.00015 | 0.016 | 0.034 | 10 | BN |
| 190 | 418 ± 29 | 20980±1635 | 0.01995±0.00278 | 0.024 | 0.129 | 10 | BN |
| 202 | 177 ± 5 | 25146±453 | 0.00704±0.00023 | 0.006 | 0.029 | 10 | BN |
| 199 | 624 ± 24 | 25842±4793 | 0.02417±0.00201 | 0.009 | 0.219 | 10 | BN |
| 212 | 16.0 ± 0.2 | N/A** | N/A** | 0.030 | 0.299 | 10 | BN |
| 214 | 140 ± 8 | 21313±846 | 0.00658±0.00042 | 0.053 | 0.151 | 10 | BN |
| 253* | 99 ± 34 | 4856±644 | 0.02050±0.00731 | 0.078 | 0.800 | 10 | BN |
| 115 | 0.6 ± 0.1 | 23622±771 | 0.00003±0.00000 | 0.002 | 2.157 | 30 | Graphite |
| 157 | 25.8 ± 2 | 9308±182 | 0.00277±0.00023 | 0.045 | 0.172 | 10 | Graphite |
| 154 | 35.9±2 | 8087±170 | 0.00444±0.00023 | 0.026 | 0.488 | 30 | Graphite |
| 149 | 137 ± 16 | 45945±338 | 0.00298±0.00036 | 0.012 | 0.182 | 10 | Graphite |
| 167* | 670 ±432 | 18609±403 | 0.03601±0.02328 | 0.079 | 0.839 | 10+30 | Graphite |

  * **Higher Mg average values from EMP analysis, ICP analysis results are accepted with a higher Mg value (>0.06 wt%).**
** **No. 212: silicate was lost during grinding.**
   **The Fe phases of some run products in Fig.1 are too small to satisfy the analysis requirement of LA-ICP-MS (>15 μm), therefore no data come from these runs. Errors are 2σ.**

small (electron beam diameter ~1 μm and sampling depth less than ~1.5 μm), and therefore, it may avoid contamination from silicate when analyzing the Fe phase. The EMP results of Gessmann and Wood (2002) showed that Mg can reach 0.17 wt% in the Fe sulfide phases at 2.5 GPa and 1600 °C. The average value of Mg in this study, from 85 analysis points in 21 samples in Fe phases by EMP analysis, was 0.06 wt%. Therefore, this value of Mg concentration is used as the upper limit for useable LA-ICP-MS data unless the run products have higher than 0.06 wt% Mg as analyzed by EMP. It is noted that the results of Dubrovinskaia et al. (2004) show that the solubility of Mg in iron can increase up to 1.99 wt % when a mixture of pure metallic Fe and Mg powder was pressurized to 20 GPa and heated to 2000°C. However, their starting material was very different from the silicate-metallic Fe system in this study. The present results indicate that Mg has a very weak increasing trend with P and T in the Fe phase. Table 3 and Fig.4 are the LA-ICP-MS analysis results after discarding data with Mg > 0.06 wt %.

According to the experimental temperature, U concentration in the Fe phase can be divided into two main groups: $T < T_{melt}$ of silicate and $T > T_{melt}$ of silicate as highlighted in Fig.4. The U concentration in the Fe phase of the first group increases slowly but stably with P. Furthermore, the U concentration in the Fe phase of the second group has a greater dependence on P than the first group. However, U concentrations of runs 212 and 253 do not follow this trend, which may be caused by their low experimental T compared with their relatively high P (Fig.1). In addition, the Fe phases of some run products in Fig.1 are too small to satisfy the minimum size requirement of LA-ICP-MS (> 15 μm), therefore, these runs did not yield any data. On the other hand, the U concentration in the silicate phases shows a scattered data set but with a generally decreasing trend with increasing pressure (Fig.7). This indicates that U gradually migrates from the silicate phase to the Fe phase with increasing pressure. As with the U concentration in the Fe phase, the partitioning coefficient $D_U$ ($U_{Fe}/U_{silicate}$) can also be divided into $T < T_{melt}$ of silicate and $T > T_{melt}$ of silicate groups (Fig.8). In both temperature groups, $D_U$ values increase with increasing pressure in the same patterns as the U concentration in the Fe phase (Fig.4). This further supports that the U solubility in the Fe phase increases with increasing P and when T is above silicate melting, more U migrates from the silicate to the Fe phase.

Fig.8 illustrates that $D_U$ values of samples from BN containers are somewhat higher than those from graphite containers. This is similar to the results of Gessmann and Wood (2002) in Fe sulfide, where the K partitioning coefficient of samples in graphite containers was one log unit lower than those from MgO and $Al_2O_3$ containers. As discussed above, C migration from the container to the Fe phase and the concomitant

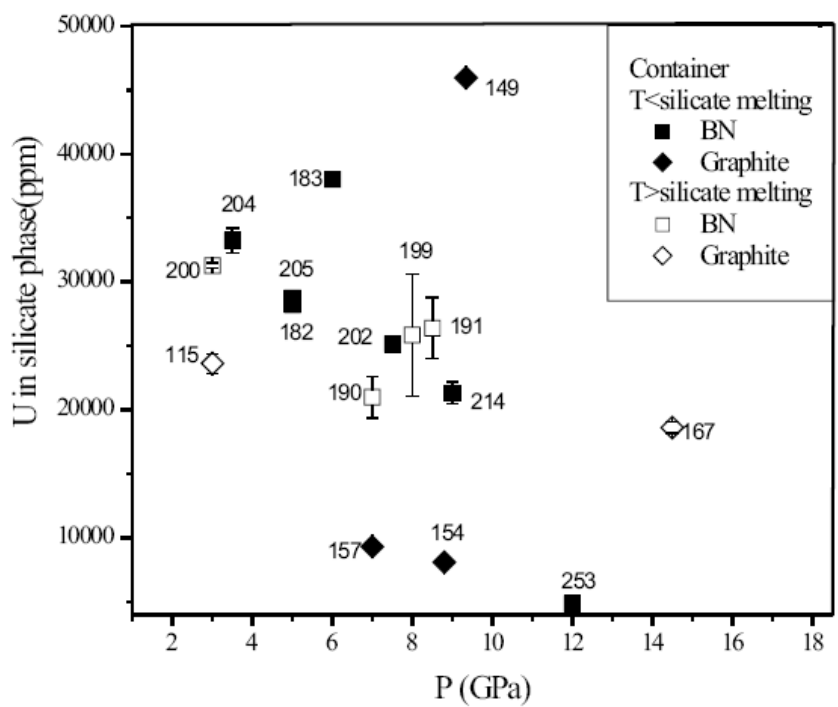

**Fig.7 U concentrations in the silicate phase as a function of pressure. The silicate phases in some of the run products in Fig.1 were lost during sample cutting and polishing. Errors are 2σ.**

volume reduction may not favor U inclusion in the Fe phase.

From the discussion above about Si in Fe phases, increasing pressure can reduce oxygen fugacity or decrease oxygen solution (Rubie et al., 2004). Previous works show that reducing conditions (low oxygen fugacity) can make U deviate from its lithophile character and allow U to enter the Fe sulfide phase (Furst et al 1982, Murrell and Burnett 1986). The present results show an increase in U concentration in the Fe phase with increasing pressure, which is consistent with an decrease in $fo_2$. Therefore, U gradually becomes less lithophile with increasing pressure.

The U concentration in the silicate phase is within the range 6,000 to 60,000 ppm. The following is proposed as a mechanism for U migration from the silicate to Fe phase. At high P and T, the U ions in uraninite ($UO_2$) may enter the silicate to substitute for $Mg^{+2}$ due to their similar atomic radii ($r_{Mg}/r_U$=1.05), or uraninite exists as small inclusions in the silicate phase. With further increase of P and T, metallic Fe may react with $UO_2$ in the silicate according to the following reaction:

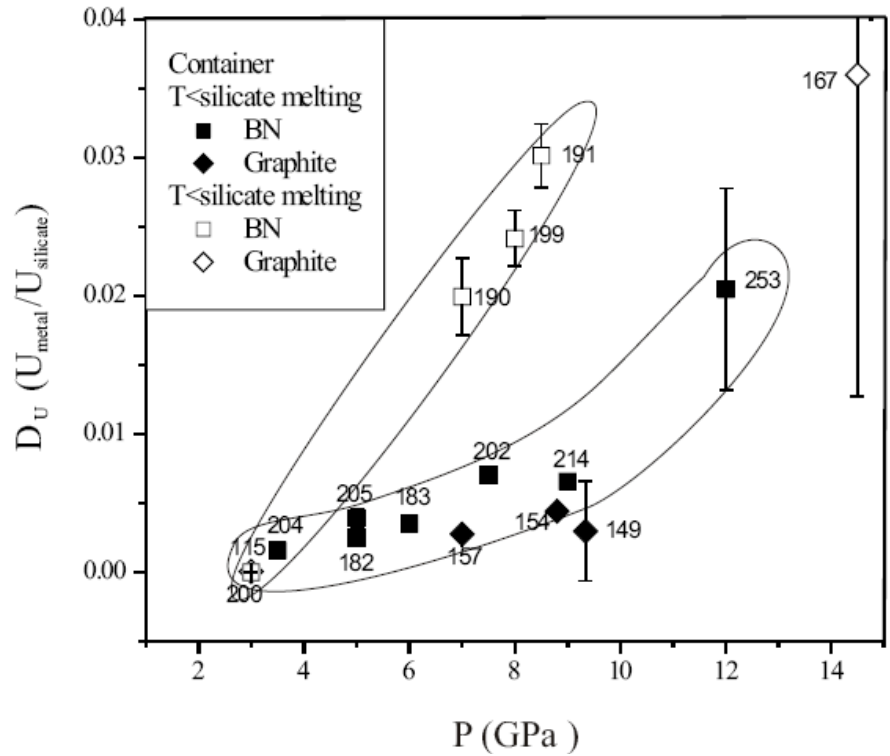

**Fig.8 Partitioning coefficient of U ($D_U$) as a function of pressure. The $D_U$ values in some of the run products in Fig.1 were absent because of the same reasons as Figs.4 and 7. Errors are 2σ.**

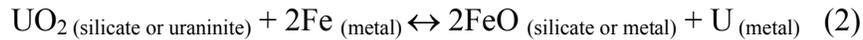

$$UO_{2 \text{ (silicate or uraninite)}} + 2Fe_{\text{ (metal)}} \leftrightarrow 2FeO_{\text{ (silicate or metal)}} + U_{\text{ (metal)}} \quad (2)$$

The fact that U in the Fe phase increases with P indicates that reaction (2) proceeds to the right with increasing P. However, it was observed that increasing temperature also causes reaction (2) to go to the right as further explained below. From Fig.4, it is seen that in runs 190, 199, 191 and 167 (at pressures 7, 8, 8.5 and 14.5 GPa, respectively), the U concentrations of the Fe phases are 4 to 6 times higher than other run products. It is noteworthy that the experimental temperatures are higher than the melting temperature of Mg-olivine (Fig.1). This suggests that when Mg-olivine melts, U ions in silicate or in uraninite included in silicate will be released and have higher mobility to enter Fe liquid, possibility as indicated in reaction (2). However, at lower pressures (e.g. runs 200 and 115), where the temperatures were also higher than (or close to) the Mg-olivine melting line, the U concentrations in the Fe phase are much lower than those of runs 167, 191, 190 or 199. This shows that in order to make reaction (2) move to right, pressure may play the dominant role, since a consequent decrease in $fo_2$ in the sample is necessary.

Similarly, increasing T also causes a reduction in $fo_2$ (Righter and Drake, 2006), which causes more U to enter the Fe phase. Therefore, the high U concentrations and large $D_U$ values of runs 167, 190, 191 and 199 appear to have caused by the relatively low $fo_2$ high created by the combination of high experimental pressures, and high temperatures.

### 3.3.2 SIMS analysis results-depth profiles

Fig.9 show the SIMS analysis results of runs 200 and 202. In general agreement with the results from LA-ICP-MS, Fig.9a and b qualitatively show that both the Fe phases contain U, and the U concentration of the higher-pressure run product (run 202, 7.5 GPa) is higher than the lower-pressure run product (run 200, 3 GPa). Fig.9c and d show the concomitant decrease in U concentration in the silicate phase of the higher-pressure run product is relative to the lower-pressure run product. This further supports the conclusion drawn from LA-ICP-MS analyses that U gradually migrates from the silicate phase to the Fe phase with increasing pressure.

The concentration of Ca is small (< 0.44wt %) in most of the Fe phases (Table 3) and no relation with U content is found, which indicates that U may alloy with Fe directly. This result is different from the lower pressure results on FeS by Murrell and Burnet (1986), who proposed that U exists in the CaS phase.

### 3.3.3 Comparison with $D_U$ values from previous studies

The $D_U$ values (0.00003 - 0.036) determined in this study overlap the wide range of $D_U$ values ($D_U$ =0.002-2) determined by Murrell and Burnett (1986) at P ≤ 1.5 GPa. The large range of $D_U$ values in their study may originate from their starting material (FeS) based on the observation that U is highly concentrated in CaS in the Fe sulfide phase (Furst et al., 1982). The $D_U$ values of the present study also overlap the range of $D_U$ values (0.001-0.012) determined by Malavergne et al. (2005) in Fe-Si and silicate systems. The decreasing $D_U$ with increasing P with the same starting composition in their study may originate from their low experimental T (at 15 GPa, their run T is 1900 $^o$C or 400 $^o$C lower than the closest P run products in this study). The conclusion that $D_U$ increases with P and T is also generally consistent with the results of Wheeler et al.

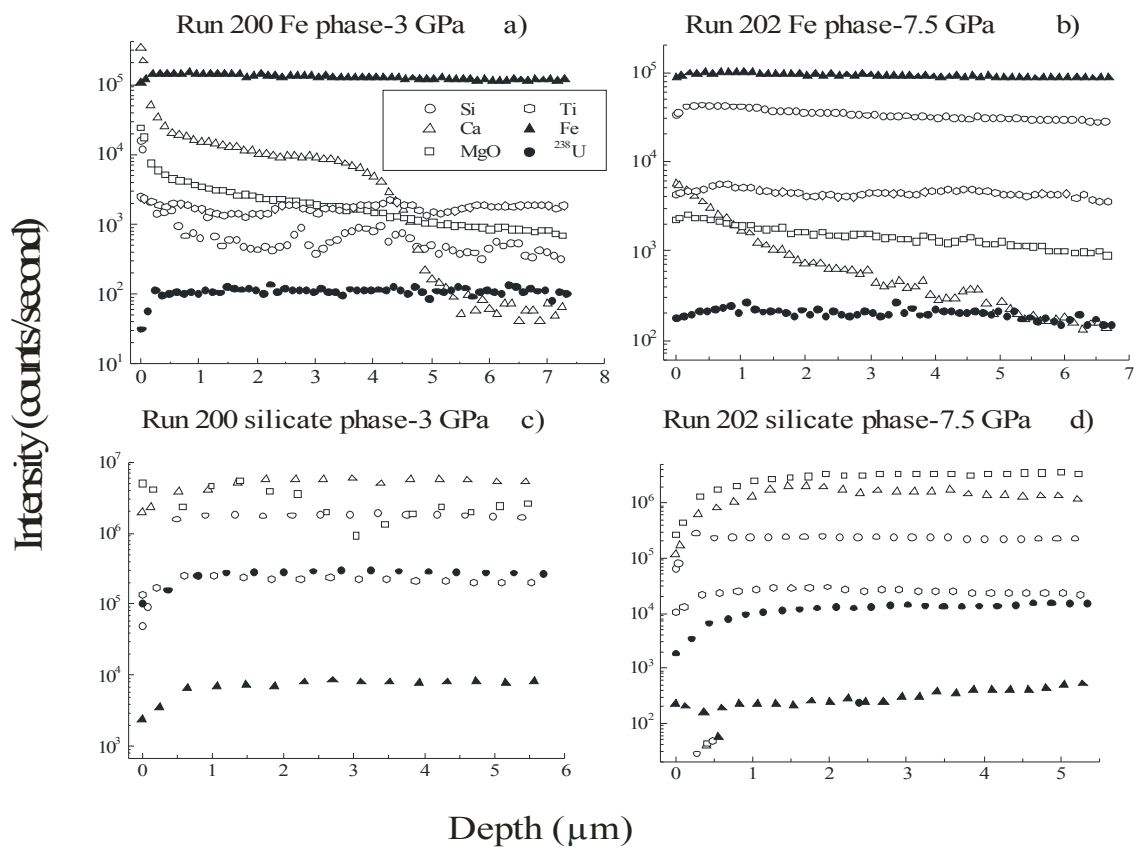

**Fig.9 Element intensity vs. depth analyzed by SIMS. For the Fe phase, a) and b); and the silicate phase, c) and d) for runs 200 and 202.**

(2004). Their $D_U$ values ($D_U < 0.001$) are smaller than those presented here, but because a complete description of their starting material is not given, a complete comparison is not possible. The small $D_U$ value may have originated from the log $fo_2 = -2$ in their experimental study (Wheeler et al., 2006), which is at the high extreme of the range of log $fo_2$ values in this study (-2 to -6) and higher than the $-3.2$ in the study of Malavergne et al. (2005), because $fo_2$ is thought to have a negative correlation with $D_U$ value in the metal phase (Furst et al., 1982; Murrell and Burnett, 1986; Malavergne et al., 2005).

## 4. APPLICATION TO THE CORES OF EARTH AND MERCURY

Currently, the generally accepted model of Earth core formation is separation from a magma ocean (Rubie et al., 2003; Righter, 2003) during Earth's accretion within a very short time span of 30 Ma (Jacobsen, 2005). Therefore, during core formation, the temperature of the magma ocean must have been higher than the melting temperature of the silicate phase. If we consider the depth of the magma ocean was approximately 700 - 800 km (Righter, 2003), then the pressure was 25-30 GPa at its base. It is therefore expected that the U partitioning coefficient at these pressures should be higher than the highest experimental $D_U$ value of 0.031 (BN container) or 0.036 (graphite container) in this study, as the U partitioning coefficient trend in Fig.8 is extrapolated to higher pressure (Fig.10).

Metallurgical studies show that when the pure elements Fe, U and Si are mixed at high T created by arc furnaces and under purified argon (reducing) atmosphere, U can substitute for Fe extensively and form a solid solution of $UFe_{10}Si_2$ (U: 27.86 wt %,

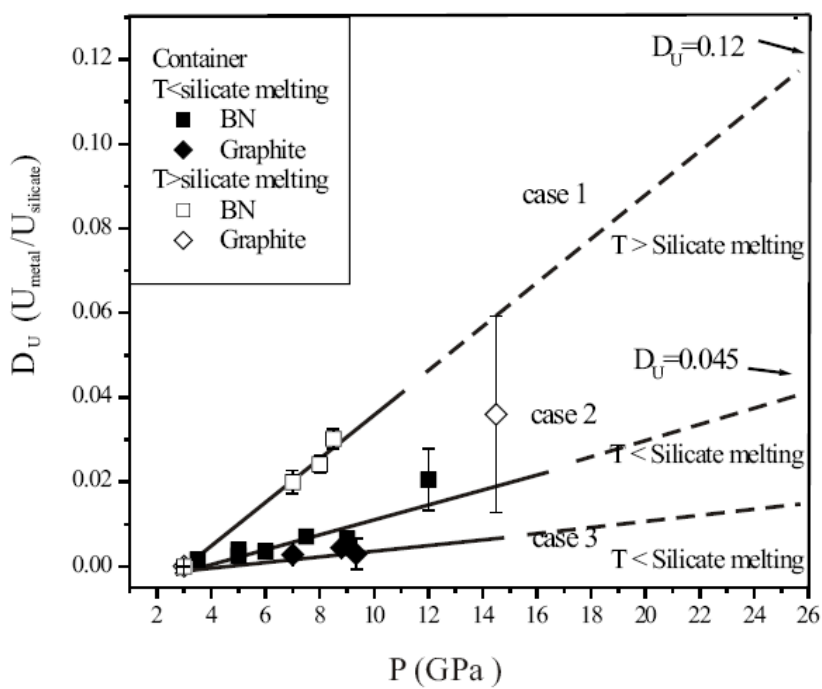

**Fig.10 Extrapolated $D_U$ values at 26 GPa, the P at the base of an intermediate magma ocean (Righter, 2003). Case 1 is based on the run products with experimental $T > T_{melt}$ of silicate phase, for the magma ocean core formation model. Case 2 is based on the run products with experimental $T < T_{melt}$ of silicate phase with the situation of T increasing with increasing P, for the percolation core formation model. Case 3 is based on the run products of $T < T_{melt}$ of silicate phase with the situation of T not increasing with increasing P, and not analogous to the mantle.**

Brandle et al., 1990). Similarly $Fe_{76.5-x}U_xCu_1Si_{13.5}B_9$ also contains up to 38 wt % U (Dusa

et al., 1996). The impact research of Yakovlev (1993) confirmed that explosive impact processes can produce reduced forms of metallic Si and Al from silicate starting material.

If the Earth's core was formed from a magma ocean, and giant impacts by smaller planetary bodies were the main mechanism to produce such a magma ocean (Rubie et al., 2003; Righter, 2003; Jacobsen, 2005), then the $UO_2$ in the starting material of the Earth could have been partially reduced to metallic U, and on the basis of present results, could have entered the core.

In the case of Mercury, Benz et al. (1988) concluded that it formed by a giant collision between a proto-Mercury and a planetesimal one-sixth its mass, which resulted in the loss of most of the silicate mantle of the planet. This scenario would have left behind an iron-rich planet and thus explain the anomalously high density of Mercury. Palme et al. (2003) pointed out that the Fe/Mg ratio of the whole Earth is at least 10% higher than the Fe/Mg ratio of the average solar system, and it cannot be explained by volatility-related fractionation in the solar nebula, because Mg and Fe have similar nebular volatilities. They further proposed a collision-erosion hypothesis to explain this paradox, namely, a late stage giant impact may have removed part of the silicate mantle and left a proportionately larger Fe core inside the Earth. Therefore, the apparent excess Fe in both Earth and Mercury strongly support a late stage giant impact origin hypothesis.

Because the terrestrial planets have lost large (Mercury) or small (Earth) fractions of their silicate mantle, when calculating the total amount of U, the lost part of the silicate mantle should be included. This means that the total U content should be somewhat higher than the traditional Cl chondritic meteorite composition model due to the fact that U is not a volatile element and was relatively fixed in location compared with other mantle compositions during these impacts. Alternatively, other researchers have pointed out that the Earth's building materials should be similar to bulk enstatite chondrite (EC) composition (Smith, 1981, Javoy, 1995, 1998), because only EC possesses the same Fe content (Javoy, 1995), oxygen (Clayton, 1993), chromium (Lugmair and Shukolyukov, 1998) and molybdenum (Dauphas et al., 2002) isotope compositions, and similar redox characteristics (Dauphas et al., 2002) to bulk Earth. According to the observation of Furst et al. (1982), the highly reducing formation conditions for EC have caused U to deviate from its lithophile behavior and enter the metal sulfide phase. Consequently, if Earth's building material is EC rather than Cl chondrites, which are usually considered to be the building material of Earth and other terrestrial planets, then there should have been more U entering the metal core rather than entering the silicate mantle and crust.

In conclusion, according to the experimental results of this study and the core formation hypotheses discussed above, there should have been a significant amount of U entering the cores of the Earth and Mercury, which if present, would contribute as an energy source to keep their cores partially liquid (Margot et al., 2004; Stanley et al., 2005) and to maintain their dipolar magnetic fields.

## 5. TOWARDS A SOLUTION OF THE CORE COOLING PARADOX

The present experimental results suggest that U could have dissolved in Earth's Fe core during its formation. From the model of timing and rate of inner core solidification, and from a consideration of the energy balance at the core-mantle boundary, Labrosse et al. (2001) and Anderson (2002) concluded that there is a need for some amount of radioactive heating in the core. Labrosse et al. (2001) argued that, in order to have an inner core age of 2.7 Ga, the concentration of U in the core should be $\geq 10$ ppb, assuming U is the only heat-producing element in the core. Olson (2006) recently suggested the need for additional core heat source(s), including radioactive heating, to explain the high temperature predicted prior to inner core formation. The amount of U that could enter the Earth's core depends on the amount of U in the silicate phase that was in contact with the Fe phase, which in turn depends on the core separation mechanism. Currently, there are two main mechanisms: magma ocean and percolation core formation hypotheses (e.g. Righter, 2003). In the magma ocean scenario, where T > silicate melting temperature (case 1, Fig.10), the largest $D_U$ value is obtained by extrapolation to pressures equal to the pressures at the base of a 700 km thick magma ocean. However, the original U concentration at the bottom of the magma ocean, which was in contact with the Fe component, is difficult to know based on current knowledge, because the Earth's major, minor, and isotopic composition is unlike any known chondrite group or mixture of chondrite groups (Righter, 2003). But as an approximation, a U concentration of silicate

Earth of 20 ppb (McDonough, 2003) can be used as the lower limit. The high density of $UO_2$ (10.95 g/cm$^3$) would cause it to sink to the bottom of the magma ocean and to increase the U concentration in the lower region that contacted the Fe liquid. Using a conservative estimate of pressure at the base of the magma ocean depth of 25-30 GPa (Rubie, et al., 2003; Righter, 2003), the $D_U$ value obtained from the current experiments at T above silicate melt to 26 GPa can be estimated by extrapolation. As shown in Fig.10, the extrapolated $D_U$ value is approximately 0.12 at such a depth. For this $D_U$ value, the U concentration of the Fe phase in such a magma ocean will reach 2.4 ppb. Moreover, the impact of small planetesimals in the late stage of Earth's accretion may have partially transformed $UO_2$ into metallic U and allow more U to enter the core as discussed above. At the same time, S may have been incorporated into Earth's core, and which would have enhanced this $D_U$ value, as indicated by the observation of Furst et al. (1982) and new experiments on Fe-10wt% S as metal components (Bao and Secco, 2006). This would result in a U concentration in the core greater than 2.4 ppb.

If Earth's core did not form from a magma ocean, then it might have formed by percolation at a temperature much lower than silicate melting (Bruhn et al., 2000), (case 2, Fig.10). In this scenario, metallic Fe gradually separated from, and percolated through, the solid silicate mantle and formed the core (Bruhn et al., 2000). Because P and T increase with depth within Earth's mantle, the experimental result of $D_U$ increasing with P and T shows that U can enter the Fe phase in Earth's mantle in increasing amounts with increasing depth, and finally enter the core. In this case, U in the whole primeval mantle or the current lower mantle could be used as a source for the core due to the P and T conditions required to enable the percolation of metallic iron (Shannon and Agee, 1998). However, based on the low P experimental results and for comparison with the magma ocean model presented above, extrapolation of $D_U$ to 26 GPa (the P at the bottom of the magma ocean) was made first. This gave a $D_U$ value of 0.045 (Fig.10). With this $D_U$ value, and again using the U concentration of the silicate Earth of 20 ppb from McDonough (2003) to calculate the U concentration of Earth's core, then the U concentration of the Earth's core is approximately 1 ppb. In addition, the $D_U$ can be extrapolated to the core-mantle boundary (135 GPa) according to the P and T range of the percolation model (Shannon and Agee, 1998), and at 135 GPa, this yields a $D_U$ value of 0.2. This extrapolation is extremely large and the $D_U$ value obtained is very speculative. Similarly, if the extrapolated $D_U$ value and the U concentration of the silicate Earth of 20 ppb from McDonough (2003) are combined to calculate the U concentration, the U concentration of the Earth's core would be 4 ppb. This result is much higher than that achieved from magma ocean core formation model. Case 3 in Fig.10 would occur only when pressure increases but temperatures decreases. This case is not applicable to the Earth's mantle, and thus it is not discussed further.

In summary, based on the Cl chondrite value of U and neglecting any S in the building material, the calculated U concentrations in the core would be > 2.4 ppb for the magma ocean and 1-4 ppb for the percolation core formation models. Therefore, U may be one of the contributors to radioactive heating in the core and possibly, in addition to K (Murthy et al., 2003), may satisfy the calculated required radiogenic heat sources of Labrosse et al. (2001) to explain an inner core age over 2.7 Ga.

For a detailed discussion of the implications for terrestrial planetary dynamics, please refer to Bao (2006).

6. CONCLUSIONS

U is soluble in the metallic Fe at pressures of 3 to 14.5 GPa and temperatures of 1900 to 2500 $^o$C with a mixture of 40 wt% Fe, ~54wt% peridotite and ~6wt% uraninite, which is similar to Earth's building material. The solubility of U ($D_U$) in the metal phase increases with increasing P and T, and when P is $\geq$ 7 GPa, and T > silicate melting, $D_U$ is 3 to 5 times higher than the $D_U$ values from low P and T run products.

A significant amount of U may have entered the cores of Earth and Mercury even if there was no S in their building materials. Uranium may be one of the radioactive heat sources that makes their cores partially liquid and provides energy to maintain their dipolar magnetic fields.

If Earth's core formed from a magma ocean with a P of 26 GPa at its bottom, and the core is composed of Fe and a small amount of Si, then more than 2.4 ppb U may have been incorporated into the core. With the same building material, if Earth's core was formed by percolation, then 1 to 4 ppb U may have been incorporated into the core. The heat energy from U may be one of the radioactive heat contributors to satisfy an inner core age of more than 2.7 Ga.

### Acknowledgments

This work was supported by a grant awarded by the National Sciences and Engineering Research Council of Canada to RAS. We thank G. Young for his partial funding support to XB. C. Cermignani, M. Liu and L. Shi for electron microprobe analysis; A. Pratt for his help with U analysis; G. Gord for SIMS analysis; D. Liu for his help in sample preparation; R. Tucker for his help in preparing high pressure components and G. Wood for his help in sample cutting and polishing; R.V. Murthy, S. Shieh, and R.A. Flemming for their comments and suggestions.